\documentclass[preprint,12pt]{elsarticle}
\biboptions{sort&compress}

\usepackage{amssymb}
\usepackage[english]{babel}
\usepackage[T1]{fontenc}
\usepackage{amsmath}
\usepackage{epstopdf}
\usepackage{geometry}
\usepackage{bm}
\usepackage{booktabs}
\usepackage{float}
\usepackage{pict2e}
\usepackage{siunitx}
\usepackage{placeins}

\DeclareMathOperator*{\sign}{sign}
\DeclareMathOperator{\rank}{rank}

\journal{Computer Physics Communications}

\begin{document}
\renewcommand{\eqref}[1]{(\ref{#1})}
\newcommand{\Eq}[1]{Eq.~(\ref{#1})}
\newcommand{\Eqs}[1]{Eqs.~(\ref{#1})}
\newcommand{\Sec}[1]{Sec.~\ref{#1}}
\newcommand{\Secs}[1]{Secs.~\ref{#1}}
\newcommand{\Ref}[1]{Ref.~\cite{#1}}
\newcommand{\Refs}[1]{Refs.~\cite{#1}}
\newcommand{\Fig}[1]{Fig.~\ref{#1}}
\newcommand{\Figs}[1]{Figs.~\ref{#1}}
\newcommand{\Tab}[1]{Table~\ref{#1}}
\newcommand{\Tabs}[1]{Tables~\ref{#1}}
\newcommand {\beq} {\begin{eqnarray}}
\newcommand {\eeq} {\end{eqnarray}}
\newcommand {\etal} {\mbox{\it et al}}
\newcommand {\e} {{\rm e}}

\begin{frontmatter}



\title{Computing an orthonormal basis of symmetric or antisymmetric hyperspherical harmonics}

\cortext[cor1]{Corresponding author.}
\author[label1]{J. Dohet-Eraly\corref{cor1}}
\ead{jdoheter@ulb.ac.be}
\author[label2,label3]{M. Viviani}
\ead{michele.viviani@pi.infn.it}
\address[label1]{Physique Quantique, and
Physique Nucl\'eaire Th\'eorique et Physique Math\'ematique, C.P. 229, Universit\'e libre de Bruxelles (ULB), B-1050 Brussels, Belgium.}
\address[label2]{Istituto Nazionale di Fisica Nucleare, Sezione di Pisa, 
Largo B. Pontecorvo 3, I-56127, Pisa, Italy}
\address[label3]{Department of Physics ``E. Fermi'', University of Pisa, 
Largo B. Pontecorvo 3, I-56127, Pisa, Italy}

\begin{abstract}
A numerical method to build an orthonormal basis of properly symmetrized hyperspherical harmonic functions is developed. As a part of it, refined algorithms for calculating the transformation coefficients between hyperspherical harmonics constructed from
different sets of Jacobi vectors are derived and
discussed. Moreover, an algorithm to directly determine
the numbers of independent symmetric hyperspherical states (in case of bosonic systems) and 
antisymmetric hyperspherical-spin-isospin states (in case of fermionic systems) entering the expansion of the $A$-body wave functions is presented.  Numerical implementations for systems made with up to five bodies are reported.
\end{abstract}




\end{frontmatter}



\section{Introduction}
The study of few-body systems is still an active area of research, with the aim of understanding the dynamics of
various atomic, molecular, nuclear, and subnuclear processes~\cite{LO13,MS15,RVV18,Pl15}. In recent years, there have been tremendous advances for solving the three- and four-body problems. However, the study of larger systems is still in progress. It is therefore of interest the development and/or refinement of sophisticated computational methods aimed to tackle this problem. 

The hyperspherical harmonics (HH) formalism is one of
the possible tool for studying few-body systems~\cite{ZB35,Av89}.
In recent years the HH technique has been applied with noticeable success to study various three- and four-body systems~\cite{KRV08,LO13}.
The HH formalism can be applied not only to describe bound-states but also for scattering processes. The interest of these studies ranges from the necessity to have a better understanding of the interaction between the constituents of these systems, the possibility to compute rates of nuclear reactions of astrophysical interest, of
various chemical reactions, for the production of energy, etc.\ The attempts to apply the formalism to larger systems face two major obstacles: 1) the difficulty to construct states with the needed symmetry under particle exchanges and 2) the large degeneracy of the basis. Both issues are addressed in this paper.

In the HH approach, the $A$-body wave function is expanded as a sum of products of an HH function and a one-dimensional function of the hyperradius $\rho$. In the case of identical particles, the hyperradius is symmetrical under any exchange of all particles. Therefore, to construct symmetrical or antisymmetrical states one has to
use HH functions with the appropriate symmetry (eventually, taking into account also spin and isospin degrees of freedom). 
A standard way for obtaining totally symmetric or antisymmetric states is to apply to the HH functions the symmetrization or antisymmetrization projectors, which involve sums over all particle permutations. The calculation of the matrix elements of the interaction between  such states is rather complex, and usually
the integrals to be evaluated require the integration
over all the $3A-3$ coordinates (in addition to the trace over the spin-isospin degrees of freedom), a task that becomes more and more difficult as $A$ increases. Moreover, in case of strong repulsive two-body interactions, many HH states have to be included in the expansion of the $A$-body wave function and the task of computing the matrix elements becomes more and more involved.

A number of different methods have been devised in order to deal with the problem of the symmetrization. In Refs.~\cite{BN97,BN98,Ba99,BLO00,BLO03,BLO10,BBS12}, 
HH basis functions belonging to well defined irreducible
representations of the $A$-body symmetric group 
are constructed in terms of those of the $(A-1)$-body system while in \Refs{Ti02,Ti04,Ti08}, antisymmetrical HH functions are obtained in terms of combinations of Slater determinants of the oscillator translation-invariant shell model. Recently, another approach has been developed where hyperspherical calculations are performed using correlated Gaussian basis functions constrained at fixed values of the hyperradius~\cite{RSI11,RB12,DG14,Su15,SV19}. Alternatively, one can use directly the non-symmetrized HH basis, exploiting the fact that the exact eigenvectors of the Hamiltonian matrix belongs to well defined irreducible representations of the symmetric group. 
This requires that, for each value of the grand angular momentum $K$, all HH functions are included in the expansion of the wave function. The difficulty in this approach is therefore connected to the identification of the states with the required symmetry between all those obtained~\cite{GKV09,GKV11a,DBL13}.

As another option, one can exploit the analytic properties of the HH functions in order to simplify the problem of the construction of symmetrical or antisymmetrical states. Let us indicate with ${\cal Y}_{[K]}(i,j,k,\ldots)$
a generic HH function for the permutation $(1,2,3\ldots)\rightarrow(i,j,k,\ldots)$ of the particles. The symbol ${[K]}$ stands for  the quantum numbers specifying
each particular HH function (for more details, see \Sec{sec2}). Explicitly, the functions
${\cal Y}_{[K]}(i,j,k,\ldots)\times \rho^K$  result to be
harmonic polynomials  of degree $K$ of the particle coordinates. Due to this property, there exist an {\it exact} relation relating the HH functions constructed with different permutations of the particles. For example, the permuted HH can be written as
\begin{equation}\label{eq:traco}
{\cal Y}_{[K]}(i,j,k,\ldots) = \sum_{[K^\prime]}
a_{[K],[K^\prime]} {\cal Y}_{[K^\prime]}(1,2,3,\ldots),
\end{equation}
where the summation is over all the quantum numbers $[K^\prime]$ with $K^\prime=K$. As the number of functions with a given $K$ value is finite, the sum in \Eq{eq:traco} is over a finite number of terms. The coefficients $a_{[K],[K^\prime]}$ are called transformation coefficients (TC). For the three-body system, they are well known and
can be easily calculated~\cite{RR70}. For a four-body  system, the analytic expression for the coefficients exists~\cite{JKS77}, but the numerical
evaluations are quite involved and therefore useful only in the case of low order HH functions. In \Ref{Vi98}, a method was proposed to compute such TC by means of a set of recurrence relations. The so obtained coefficients were extensively used for calculating the properties of bound and scattering states of $A=4$ systems~\citep{VKR05,KRV08}.

A number of other alternative methods (see Ref.~\cite{Vi98} and references therein)  has been developed to
calculate the coefficients of \Eq{eq:traco}. For example, in ref.~\cite{Ef95}, the TC are determined by requiring
that they explicitly verify \Eq{eq:traco} for a number of spatial configurations, usually generated in a random way. The TC are then simply obtained by solving a system of linear equations. However, as $K$ (and $A$) increases, the size of this linear system becomes very large and this method becomes impractical. In the present paper, we discuss an improvement of such a method with the aim of reducing the size of the linear system to be solved by using the known properties of the HH functions under parity. In addition, we present the extension to a generic $A$-body system of the recurrence relations discussed in Ref.~\cite{Vi98} for computing the TC. 
This method is particularly adapted for HH functions having small values of the relative orbital angular momenta, which are the ones giving the most important contribution for bound states or low-energy processes~\cite{KRV08}.

By projecting a set of linearly independent hyperspherical harmonics on the symmetric or antisymmetric space, one usually gets a set of linearly dependent functions. When this set is highly redundant, an important fraction of the computational time is used to build symmetric or antisymmetric HH functions which are eventually discarded. In this paper, we devise a method to determine a linearly independent set of properly symmetrized HH without evaluating any transformation coefficient. After having actually computed this basis set, it can be made orthonormal by standard algebra techniques. 

In \Sec{sec2}, the hyperspherical harmonics formalism is outlined. The algorithms devised for building an orthonormal basis of symmetric or antisymmetric HH functions are presented in \Sec{sec3} and applied in \Sec{sec4}. Conclusions and prospects are provided in \Sec{sec:conc}.
\section{Hyperspherical harmonics formalism}\label{sec2}
We briefly present here the hyperspherical harmonics formalism, which enables us to introduce our notations. More details can be found, for instance, in \Ref{KRV08}. For a system of $A$ particles with equal masses, one possible definition of the $N=A-1$ internal Jacobi coordinates is 
\begin{equation}\label{defJac}
\bm{x}_{N-j+1}=\sqrt{\frac{2 j}{j+1}} \left( \bm{r}_{j+1} -\frac{1}{j}\sum^{j}_{i=1} \bm{r}_{i}\right),
\end{equation}
where $\bm{r}_j$ is the coordinate of nucleon $j$ and $j=1,\ldots,N$. Other choices of Jacobi coordinates are discussed in \Sec{OtherJacob}. From the Jacobi coordinates, the hyperspherical ones $(\rho,\Omega)$ can be defined. The hyperradius $\rho$ is given by
\begin{equation}\label{rhodef}
\rho=\sqrt{\sum^N_{i=1} x^2_i}
=\sqrt{\frac{2}{A}\sum^A_{j>i=1} (\bm{r}_i-\bm{r}_j)^2}.
\end{equation}
It is symmetric with respect to particle exchanges and does not depend on the specific choice of Jacobi coordinates. The hyperangular coordinates are given by
\begin{equation}
\Omega=(\bm{\hat{x}}_1,\ldots,\bm{\hat{x}}_N,\phi_2,\ldots,\phi_N),
\end{equation}
where $\bm{\hat{x}}_i=(\theta_i,\varphi_i)$ is the angular part of the spherical coordinates of $\bm{x}_i$ with $i=1,\ldots,N$.
Using the standard definition, the hyperangles $(\phi_2,\ldots,\phi_N)$ are given by 
\begin{equation}\label{hyperang}
\cos\phi_i=\dfrac{x_i}{\sqrt{\sum^i_{j=1} x^2_j}}
\end{equation}
with $i=2,\ldots,N$. The hyperangular part of the volume element reads
\begin{equation}
d\Omega=\left[ \prod^N_{j=1} \sin\theta_j d\theta_j d\varphi_j \right] \prod^N_{j=2} (\cos\phi_j)^2 (\sin\phi_j)^{3 j-4} d\phi_j.
\end{equation}
The hyperspherical harmonics are defined by
\begin{eqnarray}\nonumber
\mathcal{Y}^{KLM}_{[l_N L_{N-1} n_N]}(\Omega)=[[\ldots[Y_{l_1}(\hat{x}_1)Y_{l_2}(\hat{x}_2)]_{L_2} \ldots Y_{l_{N-1}}(\hat{x}_{N-1})]_{L_{N-1}} Y_{l_N}(\hat{x}_N)]_{LM} \\\label{HHcoupl}
\times\prod^N_{j=2}
\mathcal{P}^{\alpha_{K_{j-1}},\beta_{l_j}}_{K_j}(\phi_j),
\end{eqnarray}
where the notation $[l_N L_{N-1} n_N]$ stands for $l_1,\ldots,l_N,L_2,\ldots,L_{N-1},n_2,\ldots,n_N$, $Y_{lm}$ is a spherical harmonic function, $K$ is the grand angular momentum, $L$ is the total orbital angular momentum, $M$ is its projection on the $z$ axis, 
\beq
K_j&=&\sum^j_{i=1} (l_i+2 n_i)
\eeq
with $n_1\equiv 0$ and $j=1,\ldots,N$, and 
\beq
\alpha_{K_{j-1}}&=&K_{j-1}+\dfrac{3}{2}j-\dfrac{5}{2},\\\beta_{l_j}&=&l_j+\dfrac{1}{2}
\eeq
with $j=2,\ldots,N$. The functions $\mathcal{P}^{\alpha_{K_{j-1}},\beta_{l_j}}_{K_j}$ read, in terms of Jacobi polynomials $P^{\alpha_{K_{j-1}},\beta_{l_j}}_{n_j}$,
\begin{equation}
\mathcal{P}^{\alpha_{K_{j-1}},\beta_{l_j}}_{K_j}(\phi_j)=
\mathcal{N}^{\alpha_{K_{j-1}},\beta_{l_j}}_{n_j} (\cos\phi_j)^{l_j} (\sin\phi_j)^{K_{j-1}} P^{\alpha_{K_{j-1}},\beta_{l_j}}_{n_j}(\cos 2\phi_j),
\end{equation}
where the normalization factors are given by
\begin{equation}
\mathcal{N}^{\alpha,\beta}_{n}=\sqrt{\frac{2 (2n+\alpha+\beta+1)\Gamma(n+1)\Gamma(n+\alpha+\beta+1)}{\Gamma(n+\alpha+1)\Gamma(n+\beta+1)}}
\end{equation}
and $\Gamma$ is the Gamma function. The hyperspherical harmonics are orthonormal and constitute a complete basis. Note that other orders of coupling of the spherical harmonics could have been chosen in \Eq{HHcoupl}. However, for a given $K$, if all possible HH functions are considered, all choices of coupling ordering are equivalent. The hyperspherical harmonics are eigenvectors of the grand angular momentum operator $\Lambda^2$, of the square of the total orbital angular momentum $\bm{L}$ and of its $z$ component, and of the parity operator $\Pi$
\beq
\Lambda^2\mathcal{Y}^{KLM}_{[K]}(\Omega)&=&
-K(K+3N-2)\mathcal{Y}^{KLM}_{[K]}(\Omega),\\
L^2\mathcal{Y}^{KLM}_{[K]}(\Omega)&=&\hbar^2 L(L+1) \mathcal{Y}^{KLM}_{[K]}(\Omega),\\
L_z \mathcal{Y}^{KLM}_{[K]}(\Omega)&=&\hbar M \mathcal{Y}^{KLM}_{[K]}(\Omega),\\
\Pi\mathcal{Y}^{KLM}_{[K]}(\Omega)
&=&(-1)^{l_{\rm sum}}\mathcal{Y}^{KLM}_{[K]}(\Omega)=(-1)^{K}\mathcal{Y}^{KLM}_{[K]}(\Omega),
\eeq
where $[l_N L_{N-1} n_N]$ has been abbreviated as $[K]$ and
\beq
l_{\rm sum}=\sum^N_{i=1} l_i. 
\eeq
\section{Orthonormal basis of symmetric or antisymmetric hyperspherical harmonics}\label{sec3}
\subsection{Symmetric and antisymmetric hyperspherical harmonics}
The wave function of a system of particles is symmetric with respect to the exchanges of identical bosons and antisymmetric with respect to the exchanges of identical fermions. To guarantee these symmetry properties in variational HH approaches, one can expand the wave function on a basis of properly symmetrized hyperspherical harmonics. To illustrate the method, two different types of systems are studied throughout this paper. 

Let us consider first a system of $A$ identical bosons with spin zero. The hyperspherical harmonics~\eqref{HHcoupl} are symmetric with respect to the exchange of bosons $1$ and $2$ if $l_N$ is even and antisymmetric if $l_N$ is odd. For other boson exchanges, they are in general neither symmetic nor antisymmetric. However, fully symmetric HH functions can be built by means of the symmetrization projector defined by
\beq\label{defS}
\mathcal{S}=\frac{1}{A!}\sum_{P} \widehat{P},
\eeq 
where the sum is over all the possible permutations of $A$ particles and $\widehat{P}$ is the operator producing the given permutation of the particles. The symmetric HH are given by
\begin{eqnarray}\label{defHHS}
\mathcal{Y}^{KLM,\mathcal{S}}_{[K]}(\Omega)=
\mathcal{S}\mathcal{Y}^{KLM}_{[K]}(\Omega)
=\frac{2}{A!}
\sum_{{\rm even}\ P} \mathcal{Y}^{KLM}_{[K]}(\Omega^p),
\end{eqnarray}
where the sum is restricted to the even permutations of $A$ bosons, $l_N$ is assumed to be even, and $\Omega^p$ designates the hyperangular coordinates $\Omega$ under permutation $P$.

As a second example, a system of $A$ nucleons is considered. Within the isospin formalism, the $A$-nucleon wave function has to be antisymmetric with respect to nucleon exchanges. The $A$-nucleon wave function contains spatial, spin, and isospin parts. Let us define the spin functions $\chi^{SM_S}_{[S]}$ with total spin $S$ and total spin projection $M_S$ by coupling the individual spin functions $\chi_{1/2,\pm 1/2}$ of each nucleon,
\begin{equation}
\chi^{SM_S}_{[S]}=[[\ldots[\chi_{1/2}(1)\chi_{1/2}(2)]_{S_2}\ldots\chi_{1/2}(N-1)]_{S_{N-1}}\chi_{1/2}(N)]_{SM_S},
\end{equation}
where $[S]$ stands for $S_2,\ldots,S_{N-1}$. The isospin functions $\xi^{TM_T}_{[T]}$ with total isospin $T$ and total isospin projection $M_T$ are defined in a similar way, where the notation $[T]$ stands for $T_2,\ldots,T_{N-1}$. Hyperspherical harmonics including spin and isospin parts are obtained as products of HH, spin, and isospin functions,
\beq\label{HHST}
\mathbb{Y}^{KLSTMM_SM_T}_{[KST]}(\Omega)=\mathcal{Y}^{KLM}_{[K]}(\Omega)\chi^{SM_S}_{[S]}\xi^{TM_T}_{[T]},
\eeq
where $[KST]$ stands for $[K][S][T]$. These functions are symmetric with respect to the exchange of nucleons $1$ and $2$ if $l_N+S_2+T_2$ is even and antisymmetric if $l_N+S_2+T_2$ is odd. Let us introduce the antisymmetrization projector defined by
\beq
\mathcal{A}=\frac{1}{A!}\sum_{P} \sign(P) \widehat{P},
\eeq 
where the sum is over all the possible permutations of $A$ nucleons and $\sign(P)$ is the signature of permutation $P$. Hypersphical harmonics respecting the Pauli principe are obtained as
\beq\nonumber
\mathbb{Y}^{KLSTMM_SM_T,\mathcal{A}}_{[KST]}(\Omega)
&=&\mathcal{A}\mathbb{Y}^{KLSTMM_SM_T}_{[KST]}(\Omega)\\\label{HHPauli}
&=& \frac{2}{A!}
\sum_{{\rm even}\ P} \mathcal{Y}^{KLM}_{[K]}(\Omega^p) (\widehat{P}\chi^{SM_S}_{[S]})(\widehat{P}\xi^{TM_T}_{[T]}),
\eeq
where the sum is restricted to the even permutations of $A$ nucleons and $l_N+S_2+T_2$ is assumed to be odd.

Whatever the type of identical particles which is considered, what needs to be evaluated for building properly symmetrized HH is the effect of a permutation on the HH functions and possibly on the spin and isospin functions. This is discussed in \Secs{secTC}-\ref{sec35}.

In contrast to the unsymmetrized HH, the symmetric and antisymmetric ones are not orthonormal. Moreover, since the spaces of symmetric or antisymmetric functions are subspaces of the arbitrary-symmetry function space, many of the symmetric or antisymmetric HH are linearly dependent, in general. A method for extracting a basis of properly symmetrized and making it orthonormal is presented in \Sec{secorth}.
\subsection{Definition of the transformation coefficients of hyperspherical harmonic functions}\label{secTC}
Let us discuss the effect of a permutation of the particles on an HH function. 
Since the HH functions constitute a basis and since the grand angular operator and the total orbital angular momentum are fully symmetric, the permuted HH functions $\mathcal{Y}^{KLM}_{[l_N L_{N-1} n_N]}(\Omega^p)$ can be written as linear combinations of unpermuted HH functions $\mathcal{Y}^{KLM}_{[l'_N L'_{N-1} n'_N]}(\Omega)$ with same $K$, $L$, and $M$ values.
The transformation coefficients (TC) of an HH function under a permutation $p$ are defined implicitly as
\begin{equation}\label{TCdefimp}
\mathcal{Y}^{KLM}_{[K]}(\Omega^p)=\sum_{[K']} a^{KL,p}_{[K];[K']} \mathcal{Y}^{KLM}_{[K']}(\Omega)
\end{equation}
or explicitly by
\begin{eqnarray}\label{TCdef}
a^{KL,p}_{[K];[K']}=\int d\Omega
\left[\mathcal{Y}^{KLM}_{[K']}(\Omega)\right]^*
\mathcal{Y}^{KLM}_{[K]}(\Omega^p).
\end{eqnarray}
They do not depend on the quantum number $M$. Since the permutation operators are unitary, the permuted HH are, as the unpermuted ones, orthonormal and therefore,
\begin{eqnarray}\label{check1}
\sum_{[K'']} a^{KL,p}_{[K];[K'']}  a^{KL,p}_{[K'];[K'']}=\delta_{[K],[K']}
\end{eqnarray}
for all sets of quantum numbers $[K]$ and $[K']$. These relations constitute a useful check when one develops a computational code to calculate the TC. They also provide an estimate of the numerical accuracy of the computed coefficients.

The TC are useful not only for building properly symmetrized HH but also for evaluating matrix elements between HH for two-, three-, or even more-body operators. In the next two sections, two methods for computing the transformation coefficients are presented.
\subsection{Computing the transformation coefficients: the projection method}\label{meth1}
Let us denote with $N_{KL}$ the number of HH functions for a given $(K,L,M)$ and therefore the number of terms in expansion~\eqref{TCdefimp}. Since relation~\eqref{TCdefimp} is valid for any value of the coordinates $\Omega$, by considering $N_{KL}$ coordinates $\Omega_1,\ldots,\Omega_{N_{KL}}$, one gets for each set of quantum numbers $[K]$ a system of $N_{KL}$ linear equations,
\begin{equation}\label{syst1}
\mathcal{Y}^{KLM}_{[K]}(\Omega^p_i)=\sum_{[K']} a^{KL,p}_{[K];[K']} \mathcal{Y}^{KLM}_{[K']}(\Omega_i)
\end{equation}
with $i=1,\ldots,N_{KL}$, where the coefficients $a^{KL,p}_{[K];[K']}$ are the $N_{KL}$ unknowns. The matrix of the linear system~\eqref{syst1}, denoted by $A_{KL}$, is independent on the considered permutation $p$ and on the quantum numbers $[K]$. If the coordinates $\Omega_1, \ldots$, $\Omega_{N_{KL}}$ are chosen randomly, the linear system~\eqref{syst1} is most probably not singular and the TC can then be obtained by solving it. This is the method proposed by Efros in \Ref{Ef95}. It is particularly simple and efficient when $N_{KL}$ is small. For high values of $N_{KL}$, i.e.\ when $K$ and/or $A$ is high, the method becomes numerically impractical. The first reason is that when the linear system \eqref{syst1} is big, for randomly selected coordinates $\Omega_1, \ldots$, $\Omega_{N_{KL}}$, equations~\eqref{syst1} are close to be linearly dependent and the matrix $A_{KL}$ is badly conditioned, in general. The second reason is that the bigger the system~\eqref{syst1} is, the longer the computational time required for solving it is. With direct methods, this time grows typically like $N^3_{KL}$. 

We propose here a way to soften strongly these difficulties. Let us focus first on the linear dependance problem. If the rows of $A_{KL}$ were exactly linearly dependent, it would mean that at least one direction of $\mathbb{R}^{n\times 1}$ is not spanned by the rows of $A_{KL}$. To reduce the probability that such a situation occurs, one can simply increase the number of rows in $A_{KL}$, i.e.\ considering more than $N_{KL}$ random coordinates $\Omega_i$. Then, one gets for each set of quantum numbers $[K]$ a system of $M_{KL}$ linear equations
\begin{equation}\label{syst2}
\mathcal{Y}^{KLM}_{[K]}(\Omega^p_i)=\sum_{[K']} a^{KL,p}_{[K];[K']} \mathcal{Y}^{KLM}_{[K']}(\Omega_i)
\end{equation}
with $i=1,\ldots,M_{KL}$, $M_{KL}=\beta N_{KL}$, and $\beta>1$. The $\beta$ factor is typically chosen between $1.2$ and $1.5$. Although the linear system~\eqref{syst2} has more equations than unknowns, it has exactly one solution. The $M_{KL}-N_{KL}$ redundant equations in \Eq{syst2} are useless in exact arithmetics but plays a crucial role in finite arithmetics to improve conditionning of the linear system and therefore, the numerical accuracy on the TC. 
The linear system~\eqref{syst2} can be solved by QR factorization~\citep{GL96}. A similar technique can be used to improve the accuracy of the method proposed in \Ref{Vi98} for evaluating the TC for HH functions belonging to the so-called triplet basis.

To address the second difficulty of Efros's method, namely to reduce the computational time, one needs to reduce the size of the system to be solved. It can be done by taking advantage of the symmetries of the spherical and hyperspherical harmonics. Let $\Pi_i$ be the parity operator for the coordinate $\bm{x}_i$. It is defined by
\beq\label{Pii}
\Pi_{i} f(\bm{x}_i)=f(-\bm{x}_i)
\eeq
for any function $f$. The effect of $\Pi_i$ on an HH is given by
\beq
\Pi_{i}\mathcal{Y}^{KLM}_{[K]}(\Omega)=(-1)^{l_i}\mathcal{Y}^{KLM}_{[K]}(\Omega)
\eeq
since $\Pi_{i} Y_{l_i m_i}(\hat{x}_i)=(-1)^{l_i} Y_{l_i m_i}(\hat{x}_i)$. From the parity operator $\Pi_i$, one can build the operators
\beq
\dfrac{1}{2}(1\pm\Pi_i),
\eeq
which are the projectors on the even (plus sign) or odd (minus sign) functions of $\bm{x}_i$. By projecting relation~\eqref{TCdefimp}, one gets the relations
\beq\label{relpar}
\dfrac{1}{2^{N-1}}(1+s_1\Pi_1)\ldots(1+s_{N-1}\Pi_{N-1})
\mathcal{Y}^{KLM}_{[K]}(\Omega^p)=\sum_{[K']}\! ' a^{KL,p}_{[K];[K']} \mathcal{Y}^{KLM}_{[K']}(\Omega),
\eeq
where $s_i=\pm 1$ for $i=1,\ldots,N-1$ and the prime symbol indicates that the sum is restricted to values of $l'_1,\ldots,l'_{N-1}$ such as
\beq
(-1)^{l'_i}=s_i
\eeq
for all $i=1,\ldots,N-1$. In other words, for $i=1,\ldots,N-1$, the sign of $s_i$ fixes the parity of $l'_i$. Note that the parity of $l'_N$ is also fixed since $(-1)^{l'_1+\ldots+l'_N}=(-1)^K$. Since each parameter $s_i$ can take two values, there are $2^{N-1}$ possible sets of values for $(s_1,\ldots,s_{N-1})$. The unique relation~\eqref{TCdefimp}, where all the TC for a given set of quantum numbers $[K]$ appear, can be replaced by $2^{N-1}$ relations involving each a subset of the coefficients $a^{KL,p}_{[K];[K']}$. Let us denote by $(s^{(j)}_1,\ldots,s^{(j)}_{N-1})$ the values of the coefficients $(s_1,\ldots,s_{N-1})$ for the $j^{\rm th}$ relation~\eqref{relpar} and by $N^{(j)}_{KL}$ the number of TC involved in this relation. The numbers $N^{(j)}_{KL}$ are related by 
\beq
\sum^{2^{N-1}}_{j=1} N^{(j)}_{KL}=N_{KL}.
\eeq 
As \Eq{TCdefimp}, relations \eqref{relpar} are valid for any set of coordinates $\Omega$. For the $j^{\rm th}$ relation~\eqref{relpar}, one can randomly choose $M^{(j)}_{KL}=\beta  N^{(j)}_{KL}$ coordinates $\Omega_i$, so that the associated TC are obtained by solving the linear system 
\beq\label{eq34}
\dfrac{1}{2^{N-1}}(1+s^{(j)}_1\Pi_1)\ldots(1+s^{(j)}_{N-1}\Pi_{N-1})
\mathcal{Y}^{KLM}_{[K]}(\Omega^p_i)=\sum_{[K']}\! ' a^{KL,p}_{[K];[K']} \mathcal{Y}^{KLM}_{[K']}(\Omega_i),
\eeq
with $i=1,\ldots,M^{(j)}_{KL}$. 

What are the advantages of \Eq{eq34} on \Eq{syst2}? For simplicity, let us answer by discussing specifically the optimal case where $N^{(j)}_{KL}$ is the same for all $j$, i.e.\ $N^{(j)}_{KL}=N_{KL}/2^{N-1}$ for $j=1,\ldots,2^{N-1}$. For high values of $N_{KL}$, the actual situation is not far from the optimal one as illustrated in \Tab{tabproj}. 
\begin{table}[ht]
\begin{center}
\begin{tabular}{c r r r}
\toprule
$K$&$N_{KL}$&$N^{\rm proj}_{KL}$&ratio\\ 
\midrule
0&1&1&1\\
2&5&2&2.5\\
4&15&6&2.5\\
6&35&11&3.2\\
8&70&22&3.2\\
10&126&36&3.5\\
12&210&60&3.5\\
14&330&90&3.7\\
16&495&135&3.7\\
18&715&190&3.8\\
20&1001&266&3.8\\
\bottomrule
\end{tabular}
\caption{Comparison for a 4-body system with $L=0$ between the number $N_{KL}$ of transformation coefficients and the maximal number of transformation coefficients $N^{\rm proj}_{KL}=\max\limits_{j=1,\ldots,2^{N-1}} N^{(j)}_{KL}$ involved in one of the projected relations~\eqref{relpar}. The ratio $N_{KL}/N^{\rm proj}_{KL}$ is also given and is to be compared with the optimal one: $2^{N-1}=4$.}
\label{tabproj}
\end{center}
\end{table}
Without the projections, the transformation coefficients are obtained by solving one linear system of $M_{KL}$ equations with $N_{KL}$ unknowns. By projecting the HH on different even and odd spaces, computing the transformation coefficients requires to solve $2^{N-1}$ linear systems of $M^{KL}/2^{N-1}$ equations, each with $N^{KL}/2^{N-1}$ unknowns. Since the cost of the QR factorization of a $m\times n$ matrix with $m\geq n$ is of order ${\cal O}(m n^2)$, one can expect, in the optimal case, a reduction of the computational time needed for the QR factorization by a factor up to about $2^{2N-2}$ by using the projection method. Once the QR decomposition of a matrix $m\times n$ is known, however, solving the associate linear system has a cost of order ${\cal O}(m n)$. For this part of the calculation, the computation time in the optimal case is thus reduced by a factor equal to $2^{N-1}$. As, in general, one needs to solve a great number of linear systems having the same matrix, the projection method reduces the computational time by a factor close to $2^{N-1}$ in practice.  
\subsection{Computing the transformation coefficients: recurrence relations}\label{meth2}
The method presented in the previous section is valid for any hyperspherical harmonics. For some restricted class of hyperspherical harmonics, targeted methods have been developed which can be more efficient~\cite{Vi98}. Because of centrifugal barriers, accurate studies of bound states and low-energy scattering can be performed by restricting the expansions of the wave functions to HH with small values of $l_{\rm sum}$. For these HH, an algorithm based on recurrence relations has been proposed and developed for 4-particle systems in \Ref{Vi98}. We extend it here to systems with an arbitrary number of particles. The principle of the method is to write the TC associated with a grand angular momentum $K\!+\!2$, namely the coefficients $a^{K+2L,p}_{[l_N L_{N-1}], n_2,\ldots, n_i+1, 0,\ldots, 0;[l'_N L'_{N-1} n'_N]}$, from the ones associated with $K$ or $K\!+\!2$ and previously computed. One starts from a particular case of relation~\eqref{TCdef}, 
\begin{eqnarray}\nonumber
a^{K+2L,p}_{[l_N L_{N-1}], n_2,\ldots, n_i+1, 0,\ldots, 0;[l'_N L'_{N-1} n'_N]}=
\\\label{TCdef_r}
\int d\Omega
\left[\mathcal{Y}^{K+2LM}_{[l'_N L'_{N-1} n'_N]}(\Omega)\right]^*
\mathcal{Y}^{K+2LM}_{[l_N L_{N-1}], n_2,\ldots, n_i+1, 0,\ldots, 0}(\Omega^p),
\end{eqnarray}
where $[l_N L_{N-1}]$ stands for $l_1,\ldots,l_N,L_2,\ldots,L_{N-1}$. 
Using the recurrence relations between Jacobi polynomials~\cite{AS65}
\begin{eqnarray}
P^{\alpha,\beta}_{n+1}(x)=[a(\alpha,\beta,n)+b(\alpha,\beta,n) x] P^{\alpha,\beta}_{n}(x)
+c(\alpha,\beta,n) P^{\alpha,\beta}_{n-1}(x),
\end{eqnarray}
where
\begin{eqnarray}
a(\alpha,\beta,n) &=& \frac{(2n+\alpha+\beta+1)(\alpha^2-\beta^2)}{(2n+2) (n+\alpha+\beta+1)(2n+\alpha+\beta)},\\
b(\alpha,\beta,n) &=& \frac{(2n+\alpha+\beta+1)(2n+\alpha+\beta+2)}{(2n+2) (n+\alpha+\beta+1)}, \\
c(\alpha,\beta,n) &=& \frac{(n+\alpha)(n+\beta)(2n+\alpha+\beta+2)}{(n+1)(n+\alpha+\beta+1)(2n+\alpha+\beta)}(\delta_{n,0}-1),
\end{eqnarray}
and $\delta_{n,0}$ is the Kronecker delta, one has
\begin{eqnarray}\nonumber
\mathcal{Y}^{K+2 LM}_{[l_N L_{N-1}], n_2, \ldots, n_i+1, 0, \ldots, 0}(\Omega^p)= \left[\prod^N_{j=i+1} (\sin\phi_{jp})^2\right] (a'+b' \cos 2\phi_{ip})
\\\label{der2}
\times \mathcal{Y}^{KLM}_{[l_N L_{N-1}], n_2, \ldots, n_i, 0, \ldots, 0}(\Omega^p) 
+c' \left[\prod^N_{j=i+1} (\sin\phi_{jp})^4\right] 
\mathcal{Y}^{K-2 LM}_{[l_N L_{N-1}], n_2, \ldots, n_i-1, 0, \ldots, 0}(\Omega^p),
\end{eqnarray}
where
\begin{eqnarray}
a' &=& a(\alpha_{K_{i-1}},\beta_{l_i},n_i) \frac{\mathcal{N}^{\alpha_{K_{i-1}},\beta_{l_i}}_{n_i+1}}{\mathcal{N}^{\alpha_{K_{i-1}},\beta_{l_i}}_{n_i}} \prod^N_{j=i+1}
\frac{\mathcal{N}^{\alpha_{K_{j-1}}+2,\beta_{l_j}}_{0}}{\mathcal{N}^{\alpha_{K_{j-1}},\beta_{l_j}}_{0}},\\
b' &=& b(\alpha_{K_{i-1}},\beta_{l_i},n_i) \frac{\mathcal{N}^{\alpha_{K_{i-1}},\beta_{l_i}}_{n_i+1}}{\mathcal{N}^{\alpha_{K_{i-1}},\beta_{l_i}}_{n_i}} \prod^N_{j=i+1}
\frac{\mathcal{N}^{\alpha_{K_{j-1}}+2,\beta_{l_j}}_{0}}{\mathcal{N}^{\alpha_{K_{j-1}},\beta_{l_j}}_{0}},\\
c' &=& c(\alpha_{K_{i-1}},\beta_{l_i},n_i) \frac{\mathcal{N}^{\alpha_{K_{i-1}},\beta_{l_i}}_{n_i+1}}{\mathcal{N}^{\alpha_{K_{i-1}},\beta_{l_i}}_{n_i-1}} \prod^N_{j=i+1}
\frac{\mathcal{N}^{\alpha_{K_{j-1}}+2,\beta_{l_j}}_{0}}{\mathcal{N}^{\alpha_{K_{j-1}}-2,\beta_{l_j}}_{0}},
\end{eqnarray}
and $\phi_{jp}$ is the hyperangle $\phi_{j}$ under permutation $p$.
For the sake of brevity, the indices whose depend $a'$, $b'$, and $c'$ are omitted.

Let us rewrite the first term of the r.h.s.\ of \Eq{der2} in terms of the Jacobi coordinates instead of the hyperangles. Using
\begin{eqnarray}
(\sin\phi_{jp})^2&=&\frac{x^2_{1p}+\ldots+x^2_{j-1 p}}{x^2_{1p}+\ldots+x^2_{jp}},\\
\cos 2\phi_{ip}&=&\frac{x^2_{i p}-x^2_{1p}-\ldots-x^2_{i-1 p}}{x^2_{1p}+\ldots+x^2_{i p}},
\end{eqnarray}
where $\bm{x}_{i p}$ is the Jacobi coordinate $\bm{x}_{i}$ under the permutation $p$, one has
\begin{eqnarray}
\prod^N_{j=i+1} (\sin\phi_{jp})^2 =\frac{x^2_{1p}+\ldots+x^2_{i p}}{\rho^2} =1-\sum^N_{j=i+1} \frac{x^2_{jp}}{\rho^2}
\end{eqnarray}
and
\begin{eqnarray}
\cos 2\phi_{ip} \prod^N_{j=i+1} (\sin\phi_{jp})^2=\frac{x^2_{i p}-x^2_{1p}-\ldots-x^2_{i-1 p}}{\rho^2}=\frac{2 x^2_{i p}}{\rho^2}-1+\sum^N_{j=i+1} \frac{x^2_{jp}}{\rho^2},
\end{eqnarray}
The first term of the r.h.s.\ of \Eq{der2} can thus be rewritten as
\begin{eqnarray}\nonumber
\left[\prod^N_{j=i+1} (\sin\phi_{jp})^2\right] (a'+b' \cos 2\phi_{ip})
\mathcal{Y}^{KLM}_{[l_N L_{N-1}], n_2, \ldots, n_i, 0, \ldots, 0}(\Omega^p)=\\ \left[
a'-b'+(b'-a')\sum^N_{j=i+1} \frac{x^2_{jp}}{\rho^2}+2b'\frac{x^2_{ip}}{\rho^2}\right]
\mathcal{Y}^{KLM}_{[l_N L_{N-1}], n_2, \ldots, n_i, 0, \ldots, 0}(\Omega^p).
\end{eqnarray}
The permuted Jacobi coordinates read, in terms of the unpermuted ones, 
\begin{eqnarray}\label{jacgam}
\bm{x}_{jp}=\sum^{N}_{k=1} \gamma^{(p)}_{jk} \bm{x}_k,
\end{eqnarray}
where the $\gamma^{(p)}_{jk}$ are real coefficients depending on the permutation $p$. The permuted HH of grand angular momentum $K$ can be written in terms of the unpermuted ones by means of the corresponding TC, supposed to be already known. 
Taking the orthogonality of the HH with different grand angular momentums into account, the contribution of the first term of the r.h.s.\ of \Eq{der2} to the integral in \Eq{TCdef_r} is
\begin{eqnarray}\nonumber
\int d\Omega
\left[\mathcal{Y}^{K+2LM}_{[l'_N L'_{N-1} n'_N]}(\Omega)\right]^*
\left[\prod^N_{j=i+1} (\sin\phi_{jp})^2\right](a'+b' \cos 2\phi_{ip})\\\nonumber
 \times
\mathcal{Y}^{KLM}_{[l_N L_{N-1}], n_2, \ldots, n_i, 0, \ldots, 0}(\Omega^p)=\sum_{[l''_N L''_{N-1} n''_N]} a^{KL,p}_{[l_N L_{N-1}], n_2,\ldots, n_i, 0,\ldots, 0;[l''_N L''_{N-1} n''_N]}
\\ \nonumber \times \sum^{N}_{k,q=1}%
\left[2 b' \gamma^{(p)}_{ik} \gamma^{(p)}_{iq}+(b'-a')
\sum^{N}_{j=i+1} \gamma^{(p)}_{jk} \gamma^{(p)}_{jq}\right]\\ \label{rhs1}
\times
\int d\Omega
\left[\mathcal{Y}^{K+2LM}_{[l'_N L'_{N-1} n'_N]}(\Omega)\right]^*
\frac{\bm{x}_k\cdot\bm{x}_q}{\rho^2}
\mathcal{Y}^{KLM}_{[l''_N L''_{N-1} n''_N]}(\Omega),
\end{eqnarray}
The integrals in \Eq{rhs1} do not depend on the permutation. They can be computed by writing $\bm{x}_k\cdot\bm{x}_q/\rho^2$ in terms of hyperspherical coordinates $\Omega$, as explained in the appendix. Many of them are zero.

The second term of the r.h.s.\ of \Eq{der2} can be written as a sum of hyperspherical harmonics whose, because of the orthogonality of the HH functions, only the hyperspherical harmonics of grand angular momentum $K\!+\!2$ have a nonzero contribution to the integral in \Eq{TCdef_r},
\begin{eqnarray}\nonumber
\left[\prod^N_{j=i+1} (\sin\phi_{jp})^4\right] 
\mathcal{Y}^{K-2 LM}_{[l_N L_{N-1}], n_2, \ldots, n_i-1, 0, \ldots, 0}(\Omega^p)
\\\label{der3}
\rightarrow \sideset{}{'}\sum_{k_1,\ldots, k_{N-i}} \eta^{(i)}_{k_1\ldots k_{N-i}}
\mathcal{Y}^{K+2 LM}_{[l_N L_{N-1}], n_2
, \ldots, n_i-1,k_1\ldots, k_{N-i}}(\Omega^p),
\end{eqnarray}
where the prime indicates that the sum is restricted to the sets $\{k_j\}_{j=1,\ldots,N-i}$ with $k_j\in\{0,1,2\}$ which verify
\begin{eqnarray}
\sum^{N-i}_{j=1} k_j=2.
\end{eqnarray}
The sum with a prime in \Eq{der3} contains $(N-i)(N-i+1)/2$ terms.
The coefficients $\eta^{(i)}_{k_1\ldots k_{N-i}}$ can be obtained by using the expansions 
\begin{eqnarray}
(\sin\phi)^2 &=\cfrac{1-\cos 2\phi}{2} &=\sum^1_{j=0} \lambda^{\alpha,\beta}_j P^{\alpha,\beta}_j(\cos 2\phi),\\
(\sin\phi)^4 &=\left(\cfrac{1-\cos 2\phi}{2}\right)^2 &=\sum^2_{j=0} \mu^{\alpha,\beta}_j P^{\alpha,\beta}_j(\cos 2\phi)
\end{eqnarray}
in \Eq{der3} and by a simple identification of common terms in both sides.
The coefficients $\lambda^{\alpha,\beta}_j$ and $\mu^{\alpha,\beta}_j$ read
\begin{eqnarray}
\lambda^{\alpha,\beta}_0 &=& \frac{\alpha+1}{\alpha+\beta+2},\\
\lambda^{\alpha,\beta}_1 &=& \frac{-1}{\alpha+\beta+2},\\
\mu^{\alpha,\beta}_0 &=& \frac{(\alpha+1)(\alpha+2)}{(\alpha+\beta+2)(\alpha+\beta+3)},\\
\mu^{\alpha,\beta}_1 &=& \frac{-2 (\alpha+2)}{(\alpha+\beta+2)(\alpha+\beta+4)},\\
\mu^{\alpha,\beta}_2 &=& \frac{2}{(\alpha+\beta+3)(\alpha+\beta+4)}.
\end{eqnarray}
The coefficients $\eta^{(i)}_{k_1\ldots k_{N-i}}$ are given recursively by
\begin{eqnarray}
\eta^{(i)}_{20\ldots0}&=&\mu^{\alpha_{K_{i}}-2,\alpha_{l_{i+1}}}_{2}
\frac{\mathcal{N}^{\alpha_{K_{i}}-2,\beta_{l_{i+1}}}_0}{\mathcal{N}^{\alpha_{K_{i}}-2,\beta_{l_{i+1}}}_2}\prod^N_{j=i+2}\frac{\mathcal{N}^{\alpha_{K_{j-1}}-2,\beta_{l_{j}}}_0}{\mathcal{N}^{\alpha_{K_{j-1}}+2,\beta_{l_{j}}}_0},
 \\ \nonumber
\eta^{(i)}_{1 \delta_{2,m} \ldots \delta_{N-i,m}}&=&\mu^{\alpha_{K_{i}}-2,\alpha_{l_{i+1}}}_{1}
\frac{\mathcal{N}^{\alpha_{K_{i}}-2,\beta_{l_{i+1}}}_0}{\mathcal{N}^{\alpha_{K_{i}}-2,\beta_{l_{i+1}}}_1}
\left[\prod^{i+m-1}_{j=i+2}
\lambda^{\alpha_{K_{j-1}},\beta_{l_j}}_0 
\frac{\mathcal{N}^{\alpha_{K_{j-1}}-2,\beta_{l_j}}_0}{\mathcal{N}^{\alpha_{K_{j-1}},\beta_{l_j}}_0}
\right]\\
&&\times
\lambda^{\alpha_{K_{i+m-1}},\beta_{l_{i+m}}}_1
\frac{\mathcal{N}^{\alpha_{K_{i+m-1}}-2,\beta_{l_{i+m}}}_0}{\mathcal{N}^{\alpha_{K_{i+m-1}},\beta_{l_{i+m}}}_1} 
\prod^N_{j=i+m+1}\frac{\mathcal{N}^{\alpha_{K_{j-1}}-2,\beta_{l_{j}}}_0}{\mathcal{N}^{\alpha_{K_{j-1}}+2,\beta_{l_{j}}}_0},
\\ 
\eta^{(i)}_{0 k_2\ldots k_{N-i}}&=&\mu^{\alpha_{K_{i}}-2,\beta_{l_{i+1}}}_{0} \eta^{(i+1)}_{ k_2\ldots k_{N-i}},
\end{eqnarray}
where $m\in\{2,\ldots,N-i\}$ and a product containing no factors is equal to one by convention.
Finally, the transformation coefficients are given from \Eqs{TCdef_r}, \eqref{der2}, \eqref{rhs1}, and \eqref{der3} by
\begin{eqnarray}\nonumber
a^{K+2L,p}_{[l_N L_{N-1}], n_2,\ldots, n_i+1, 0,\ldots, 0;[l'_N L'_{N-1} n'_N]}=\sum^{N}_{k,q=1}\left[2 b' \gamma^{(p)}_{ik} \gamma^{(p)}_{iq}+(b'-a')
\sum^{N}_{j=i+1} \gamma^{(p)}_{jk} \gamma^{(p)}_{jq}\right]\\ \nonumber
\times\sum_{[l''_N L''_{N-1} n''_N]} a^{KL,p}_{[l_N L_{N-1}],n_2,\ldots,n_i,0,\ldots,0;[l''_N L''_{N-1} n''_N]}
\\ \nonumber
\times \int d\Omega
\left[\mathcal{Y}^{K+2LM}_{[l'_N L'_{N-1} n'_N]}(\Omega)\right]^* \frac{\bm{x}_k\cdot\bm{x}_q}{\rho^2}
\mathcal{Y}^{KLM}_{[l''_N L''_{N-1} n''_N]}(\Omega)\\ \label{TCrec}
+c' \sideset{}{'}\sum_{k_1,\ldots k_{N-i}} \eta^{(i)}_{k_1\ldots k_{N-i}}
a^{K+2L,p}_{[l_N L_{N-1}], n_2,\ldots,n_i-1, k_1,\ldots,k_{N-i};[l'_N L'_{N-1} n'_N]}.
\end{eqnarray}
This relation is a generalization of the diverse recurrence relations given in \Ref{Vi98} for a 4-body system. It enables the computation of the TC of HH functions belonging to the same channel, i.e.\ having a same set of quantum numbers $[l_N L_{N-1}L]$. This property allows to choose a different maximal $K$-value for each channel, which is advantageous since the convergence rate of the HH expansion with respect to the considered maximal $K$-value strongly depends on the channel~\cite{VKR05}. 

For starting the recurrence relations~\eqref{TCrec}, one needs first to compute the TC corresponding to the HH functions with $n_2=\ldots=n_N=0$. These HH functions have a grand angular momentum $K=l_{\rm sum}$, which can be kept relatively small in bound-state and low-energy scattering calculations. For these small values of $K$, the TC can be obtained efficiently and accurately by the projection method described in the previous section.
\subsection{Transformation coefficients of the spin and isospin functions}\label{sec35}
In this section, we discuss the effect of a permutation of the particles on a spin function. The case of an isospin function is similar. Since the spin functions constitute a basis of the spin space and that total spin operator is fully symmetric, the permuted spin functions can be written as linear combinations of unpermuted ones with same total spin $S$ and spin projection $M_S$,
\begin{equation}\label{spinTCdefimp}
\widehat{P}\chi^{SM_S}_{[S]}=\sum_{[S']} a^{S,p}_{[S];[S']} \chi^{SM_S}_{[S']}.
\end{equation}
The coefficients $a^{S,p}_{[S];[S']}$ are called the spin transformation coefficients. They do not depend on $M_S$. They are explicitly given by
\beq
a^{S,p}_{[S];[S']}=\langle \chi^{SM_S}_{[S']}|\widehat{P}\chi^{SM_S}_{[S]}\rangle.
\eeq
They are easily evaluated by uncoupling the spin functions $\chi^{SM_S}_{[S]}$ and $\chi^{SM_S}_{[S']}$. Indeed, when $\chi^{SM_S}_{[S]}$ and $\chi^{SM_S}_{[S']}$ are expressed in terms of the individual nucleon spin functions, the computation of $P\chi^{SM_S}_{[S]}$ and subsequently of the overlap $\langle \chi^{SM_S}_{[S']}|P\chi^{SM_S}_{[S]}\rangle$ is trivial.
\subsection{Other choices of Jacobi coordinates}\label{OtherJacob}
Up to this point, only the particular choice~\eqref{defJac} of Jacobi coordinates has been considered. It corresponds to the clustering $N+1$.  For $N>3$, however, other clustering exist. For a 4-body system for instance, they are $3+1$ and $2+2$. For a given $(KLM)$, all HH constructed from one particular set of Jacobi coordinates form a complete basis and therefore it seems there is no need for considering other sets of Jacobi coordinates. However, except for small $K$ values, only a subset of the HH functions is usually considered in HH expansions for reducing the computational effort. If only HH based on a single set of coordinates and therefore on a particular clustering are taken into account, the configurations where a different clustering of the system is important are generally badly described. To avoid this, one can consider HH functions built from different Jacobi vectors in the HH expansion of the wave functions.
Even when, for each partial wave $(KLM)$, all HH are taken into account, working with different Jacobi vectors can be useful for evaluating spectroscopic factors and studying transfer reactions, for instance.

Both methods for computing the transformation coefficients, presented in \Secs{meth1} and \ref{meth2}, can be easily adapted to the expansion of a permuted HH in terms of a linear combination of unpermuted HH based on a different set of Jacobi coordinates. 
For instance, it is sufficient in \Eqs{syst2} and \eqref{eq34} to compute the HH on the l.h.s.\ with the selected set of Jacobi vectors. Regarding the recursion method, the only change is in the definition of the coefficients $\gamma^{(p)}_{jk}$ introduced in \Eq{jacgam} relating the Jacobi vectors of different permutations and (in general) sets.
\subsection{Extracting and orthonormalizing a set of linearly independent symmetric or antisymmetric hyperspherical harmonics}\label{secorth}
Let us consider first a system of $A$ identical bosons with spin zero. The symmetrized hyperspherical harmonics $\mathcal{Y}^{KLM,\mathcal{S}}_{[l_N L_{N-1} n_N]}$ can be written from \Eqs{defHHS} and \eqref{TCdefimp} in terms of the unsymmetrized ones as
\begin{eqnarray}\label{HHSym}
\mathcal{Y}^{KLM,\mathcal{S}}_{[K]}(\Omega)=
\sum_{[K']}  a^{KL,\mathcal{S}}_{[K];[K']}\mathcal{Y}^{KLM}_{[K']}(\Omega),
\end{eqnarray}
where
\begin{eqnarray}\label{defaS}
a^{KL,\mathcal{S}}_{[K];[K']}=\frac{2}{A!}
\sum_{{\rm even}\ P} 
a^{KL,p}_{[K];[K']}.
\end{eqnarray} 
Let us note that, since the symmetrization operator~\eqref{defS} is Hermitian and is a projector, one has the following equalities
\begin{eqnarray}\label{oSS}
\langle \mathcal{Y}^{KLM,\mathcal{S}}_{[K]} | \mathcal{Y}^{KLM,\mathcal{S}}_{[K']}\rangle=\langle \mathcal{Y}^{KLM}_{[K]} | \mathcal{Y}^{KLM,\mathcal{S}}_{[K']}\rangle=
\langle \mathcal{Y}^{KLM,\mathcal{S}}_{[K]} | \mathcal{Y}^{KLM}_{[K']}\rangle
\end{eqnarray} 
and therefore,
\begin{eqnarray}\label{check2}
\sum_{[K'']}  a^{KL,\mathcal{S}}_{[K];[K'']} a^{KL,\mathcal{S}}_{[K'];[K'']}=a^{KL,\mathcal{S}}_{[K];[K']}= a^{KL,\mathcal{S}}_{[K'];[K]}
\end{eqnarray}
for all possible sets of quantum numbers $[K]$ and $[K']$. These relations constitute a useful check on the validity and on the numerical accuracy of the coefficients~\eqref{defaS}.

To extract a basis of symmetric HH, a straightforward way is the following. For a given $(KLM)$, there are $N_{KL}$ orthonormal hyperspherical harmonics. Their projections on the symmetric space are, in general, linearly dependent. For obtaining a basis of symmetric HH, one can for instance build a $N_{KL} \times N_{KL}$ matrix $A^{\mathcal{S}}_{KL}$ formed by the coefficients $a^{KL,\mathcal{S}}_{[K];[K']}$, each row of $A^{\mathcal{S}}_{KL}$ corresponding to the coefficients of the expansion of one particular symmetric HH in the unsymmetrized HH basis. A basis of symmetric HH is then obtained by extracting, from the matrix $A^{\mathcal{S}}_{KL}$, $\rank(A^{\mathcal{S}}_{KL})$ rows linearly independent. 

To reduce the effort, instead of considering all the $N_{KL}$ symmetric HH, one can select only a subset of them, namely all $\mathcal{Y}^{KLM,\mathcal{S}}_{[K]}$ with $l_{\rm sum}$ smaller than a given $l_{\rm max}$. Because of the centrifugal barriers, the contribution of these states should be dominant in bound-state or low-energy scattering calculations. Moreover, in most cases, beyond some values of $l_{\rm max}$ smaller or even much smaller than $K$, all symmetric HH are redundant, anyway.  This enables an important reduction of the number of rows which needs to be considered in the matrix $A^{\mathcal{S}}_{KL}$. 

This simple method has two important drawbacks. First, the matrix $A^{\mathcal{S}}_{KL}$ becomes quickly big as $K$ increases and secondly, it requires to compute many symmetric HH, by evaluating a lot of TC, which are finally discarded and thus of no use.   

We propose here a more economical method which enables the determination of a symmetric HH basis without expanding the symmetric HH functions on the unsymmetrized HH basis and therefore, without computing transformation coefficients. The method is based on the following fact: if $n$ functions $f_1, \ldots,$ $f_n$ are linearly dependent, then the vectors 
\beq
(f_1(x_1),\ldots,f_1(x_m)), \ldots, (f_n(x_1),\ldots,f_n(x_m))
\eeq
with $m>n$ are linearly dependent for any choice of $(x_1,\ldots,x_m)$. From this fact, an algorithm for getting a basis set of symmetric HH can be devised. First, for a given $(KLM)$, one evaluates all of the symmetric HH by means of \Eq{defHHS} at $m$ random coordinates $\Omega$. The number $m$ needs to be bigger than the maximal number, denoted by $n_{KL}$, of linearly independent symmetric HH for the considered $(KLM)$. The number $n_{KL}$ is bounded by $N_{KL}$. However, a lower value for an upper bound of $n_{KL}$ can usually be guessed from the values obtained from smaller grand angular momentum or can be obtained by a trial-and-error process. Therefore, the value chosen for $m$ can be generally much smaller than $N_{KL}$. A maximal set of linearly independent vectors can be extracted from the $N_{KL}$ vectors 
\beq
\left(\mathcal{Y}^{KLM,\mathcal{S}}_{[K^{(1)}]}(\Omega_1),\ldots,\mathcal{Y}^{KLM,\mathcal{S}}_{[K^{(1)}]}(\Omega_m)\right),\ldots,\left(\mathcal{Y}^{KLM,\mathcal{S}}_{[K^{(N_{KL})}]}(\Omega_1),\ldots,\mathcal{Y}^{KLM,\mathcal{S}}_{[K^{(N_{KL}})]}(\Omega_m)\right)
\eeq
via Gauss elimination or the modified Gram-Schmidt algorithm, for example. 
Again, the numerical effort can be strongly reduced by considering successively subsets of symmetric HH with $l_{\rm sum}\le l_{\rm max}$ for increasing values of $l_{\rm max}$ instead of all of the $N_{KL}$ symmetric HH at once. Let us note that this method can also be used if HH functions built from different sets of Jacobi coordinates are considered simultaneously. 

This method can be easily adapted to the case of an $A$-nucleon system. It only requires to replace the symmetric HH functions, $\mathcal{Y}^{KLM,\mathcal{S}}_{[K]}$, by the antisymmetric ones, $\mathbb{Y}^{KLSTMM_SM_T,\mathcal{A}}_{[KST]}$, projected on some arbitrary spin and isospin functions with spin and spin projection $(S,M_S)$ and isospin and isospin projection $(T,M_T)$. More explicitly, it means to consider instead of the symmetric HH functions
\beq
\frac{2}{A!}
\sum_{{\rm even}\ P} \mathcal{Y}^{KLM}_{[K]}(\Omega^p),
\eeq
with even $l_N$, the functions 
\beq
\frac{2}{A!}
\sum_{{\rm even}\ P} \mathcal{Y}^{KLM}_{[K]}(\Omega^p) \langle \chi^{SM_S}_{\rm ref}| P\chi^{SM_S}_{[S]}\rangle
\langle \xi^{TM_T}_{\rm ref}| P\xi^{TM_T}_{[T]}
\rangle,
\eeq
where $l_N+S_2+T_2$ is odd and the reference spin and isospin states $\chi^{SM_S}_{\rm ref}$ and $\xi^{TM_T}_{\rm ref}$ can be chosen as 
\beq
\chi^{SM_S}_{\rm ref}=\sum_{[S]} c_{[S]} \chi^{SM_S}_{[S]} 
\eeq
and
\beq
\xi^{TM_T}_{\rm ref}=\sum_{[T]} d_{[T]} \xi^{TM_T}_{[T]},
\eeq
the coefficients $c_{[S]}$ and $d_{[T]}$ being selected randomly.

Once a set of linearly independent symmetric or antisymmetric HH functions is known, it can be useful for further applications to make it orthonormal. For the sake of brevity, we discuss the orthonormalization process only in the symmetric case but the antisymmetric case can be treated in a similar way.  For orthonormalizing a set of functions, several standard algebraic methods exist~\cite{NumRecFor} like the singular value decomposition technique or the Gram-Schmidt process, for instance. The advantage of the latter method is that it can be based on the knowledge of the overlap matrix $O$ defined by
\beq
O_{ij}=\langle\mathcal{Y}^{KLM,\mathcal{S}}_{[K^{(i)}]}|
\mathcal{Y}^{KLM,\mathcal{S}}_{[K^{(j)}]}\rangle
\eeq
for $i,j=1,\ldots,N_\mathcal{S}$ and where $\left\{\mathcal{Y}^{KLM,\mathcal{S}}_{[K^{(i)}]}\right\}_{i=1,\ldots,N_\mathcal{S}}$ is the considered set of $N_\mathcal{S}$ linearly independent symmetric HH functions. 
When the symmetric and non-symmetrized HH are built from the same Jacobi coordinates, making use of \Eq{oSS}, the overlap matrix can be obtained without any computation as 
\beq\label{overcomp2}
O_{ij}=a^{KL,\mathcal{S}}_{[K^{(j)}];[K^{(i)}]}
\eeq
for all $i,j=1,\ldots,N_\mathcal{S}$. This fact, which seems to be overlooked in \Ref{Vi98}, leads to a fast orthonormalization process.

To conclude, let us note, that if the set $\left\{\mathcal{Y}^{KLM,\mathcal{S}}_{[K^{(i)}]}\right\}_{i=1,\ldots,N_\mathcal{S}}$ constitutes a basis of the symmetric HH functions with a given $(K,L,M)$, then all symmetric HH with the same quantum numbers $(K,L,M)$ can be written in this basis as
\begin{equation}\label{exp1}
\mathcal{Y}^{KLM,\mathcal{S}}_{[K^{(j)}]}=\sum^{N_\mathcal{S}}_{i,k=1} 
a^{KL,\mathcal{S}}_{[K^{(k)}];[K^{(j)}]} (O^{-1})_{ik}
 \mathcal{Y}^{KLM,\mathcal{S}}_{[K^{(i)}]},
\end{equation}
where $j=1,\ldots,N_{KL}$. From this relation, one can directly deduce the expansion coefficients in the unsymmetrized HH basis of any symmetric HH with the same $(K,L,M)$ as  
\begin{equation}\label{exp2}
a^{KL,\mathcal{S}}_{[K^{(j)}];[K^{(l)}]}=
\sum^{N_\mathcal{S}}_{i,k=1} 
a^{KL,\mathcal{S}}_{[K^{(k)}];[K^{(j)}]}
\, (O^{-1})_{ik}
\ a^{KL,\mathcal{S}}_{[K^{(i)}];[K^{(l)}]},
\end{equation}
where $j,l=1,\ldots,N_{KL}$. Relations similar to \Eqs{exp1} and \eqref{exp2} can easily be derived for the antisymmetric case.
\section{Numerical applications}\label{sec4}
All computation are performed in quadruple precision arithmetic.
\subsection{Computation of an orthonormal basis of symmetric hyperspherical harmonics}
As an illustration, the methods developed in this paper are applied to the computation of an orthonormal basis of symmetric hyperspherical harmonics for a system of five identical bosons with total orbital angular momentum $L=0$ and grand angular momentum $K=4$. The method presented in \Sec{secorth} enables one to determine that
\begin{equation}\label{bas}
\left(\mathcal{Y}^{KLM,\mathcal{S}}_{[K^{(1)}]},\mathcal{Y}^{KLM,\mathcal{S}}_{[K^{(2)}]}\right)
\end{equation}
with 
\begin{equation}
[K^{(1)}]=0,0,0,0,0,0,0,0,2 {\rm\ \ and\ \ } [K^{(2)}]=0,0,0,0,0,0,0,1,1
\end{equation}
is a basis of the space of the symmetric HH functions with $(K,L)=(4,0)$, namely $N_{4\, 0}=2$. 
The expansion coefficients of these symmetric hyperspherical harmonics in terms of the non-symmetrized ones have been computed by \Eq{defaS} and the recurrence method of \Sec{meth2}. They are reported with at most eighteen significant digits in \Tab{tabcoef} while, as specified before, we work in quadruple precision, i.e.\ the TC are computed with more than 30 digits. 
\begin{table}[ht]
\begin{center}
\begin{tabular}{c c c c c c c c c c c}
\toprule
$l'_1$ & $l'_2$ & $l'_3$ & $l'_4$ & $L'_2$ & $L'_3$ & $n'_2$ & $n'_3$ & $n'_4$ & $a^{4\,0,\mathcal{S}}_{[K^{(1)}];[K']}$ & $a^{4\,0,\mathcal{S}}_{[K^{(2)}];[K']}$ \\
\midrule
 0 & 0 & 0 & 0 & 0 & 0 & 0 & 0 & 2 &  0.109090909090909091 &  0.031095567212508499 \\
  0 & 0 & 0 & 0 & 0 & 0 & 0 & 1 & 1 &  0.031095567212508499 &  0.034188034188034188 \\
 0 & 0 & 0 & 0 & 0 & 0 & 0 & 2 & 0 &  0.122854921183435809 &  0.048573011045975365 \\
  0 & 0 & 0 & 0 & 0 & 0 & 1 & 0 & 1 &  0.008976517050372941 &  0.009869235370762834 \\
  0 & 0 & 0 & 0 & 0 & 0 & 1 & 1 & 0 &  0.011354495733805963 &  0.012483705014362760 \\
 0 & 0 & 0 & 0 & 0 & 0 & 2 & 0 & 0 &  0.130267789455785921 &  0.063654655470171812 \\
  0 & 1 & 1 & 0 & 1 & 0 & 0 & 0 & 1 & -0.063473560777553725 & -0.069786032557925302 \\
  0 & 1 & 1 & 0 & 1 & 0 & 0 & 1 & 0 &  0.034765901043041327 &  0.038223384230765636 \\
  0 & 1 & 1 & 0 & 1 & 0 & 1 & 0 & 0 &  0.000000000000000000 &  0.000000000000000000 \\
 1 & 0 & 1 & 0 & 1 & 0 & 0 & 0 & 1 & -0.049166408763189975 & -0.054056028378946655 \\
  1 & 0 & 1 & 0 & 1 & 0 & 0 & 1 & 0 &  0.026929551151118824 &  0.029607706112288501 \\
  1 & 0 & 1 & 0 & 1 & 0 & 1 & 0 & 0 &  0.000000000000000000 &  0.000000000000000000 \\
  1 & 1 & 0 & 0 & 0 & 0 & 0 & 0 & 1 & -0.034765901043041327 & -0.038223384230765636 \\
  1 & 1 & 0 & 0 & 0 & 0 & 0 & 1 & 0 & -0.043975772881613658 & -0.048349181619592993 \\
  1 & 1 & 0 & 0 & 0 & 0 & 1 & 0 & 0 &  0.034328576702962482 &  0.037742567804819858 \\
  0 & 0 & 2 & 2 & 0 & 2 & 0 & 0 & 0 &  0.112582845582004916 &  0.011252659634262962 \\
  0 & 1 & 1 & 2 & 1 & 2 & 0 & 0 & 0 &  0.051174020719093144 &  0.056263298171314812 \\
  0 & 2 & 0 & 2 & 2 & 2 & 0 & 0 & 0 &  0.086995835222458344 & -0.016878989451394444 \\
  0 & 2 & 2 & 0 & 2 & 0 & 0 & 0 & 0 &  0.086995835222458344 & -0.016878989451394444 \\
  1 & 0 & 1 & 2 & 1 & 2 & 0 & 0 & 0 &  0.039639226000704209 &  0.043581363364040896 \\
  1 & 1 & 0 & 2 & 2 & 2 & 0 & 0 & 0 &  0.028029165506084056 &  0.030816677568068285 \\
  1 & 1 & 2 & 0 & 2 & 0 & 0 & 0 & 0 &  0.028029165506084056 &  0.030816677568068285 \\
  1 & 2 & 1 & 0 & 1 & 0 & 0 & 0 & 0 &  0.000000000000000000 &  0.000000000000000000 \\
  2 & 0 & 0 & 2 & 2 & 2 & 0 & 0 & 0 &  0.076761031078639715 & -0.028131649085657406 \\
  2 & 0 & 2 & 0 & 2 & 0 & 0 & 0 & 0 &  0.076761031078639715 & -0.028131649085657406 \\
  2 & 1 & 1 & 0 & 1 & 0 & 0 & 0 & 0 &  0.000000000000000000 &  0.000000000000000000 \\
  2 & 2 & 0 & 0 & 0 & 0 & 0 & 0 & 0 &  0.076761031078639715 & -0.028131649085657406 \\
\bottomrule
\end{tabular}
\caption{The expansion coefficients $a^{KL,\mathcal{S}}_{[K^{(1)}];[K']}$ and $a^{KL,\mathcal{S}}_{[K^{(2)}];[K']}$ of the $5$-body symmetric hyperspherical harmonics $\mathcal{Y}^{KLM,\mathcal{S}}_{[K^{(1)}]}$ and $\mathcal{Y}^{KLM,\mathcal{S}}_{[K^{(2)}]}$ in terms of the non-symmetrized HH for $(K,L)=(4,0)$, $[K^{(1)}]=0,0,0,0,0,0,0,0,2$, and $[K^{(2)}]=0,0,0,0,0,0,0,1,1$. The notation $[K']$ stands for the quantum numbers $l'_1$, $l'_2$, $l'_3$, $l'_4$, $L'_2$, $L'_3$, $n'_2$, $n'_3$, and $n'_4$.}
\label{tabcoef}
\end{center}
\end{table}
The accuracy of the expansion coefficients can be probed by checking the equalities~\eqref{check2}, which are verified within a numerical uncertainty of $10^{-33}$. Any other symmetric HH with $(K,L)=(4,0)$ can be also expanded in terms of the unsymmetrized HH basis by means of \Eq{defaS} and the methods of \Secs{meth1} and \ref{meth2}. Alternatively, the expansion coefficients~\eqref{defaS} can be obtained from the coefficients given in \Tab{tabcoef} by using \Eq{exp2}. 
The consistency of both approached has been checked numerically.

An orthonormal basis $\left(\mathcal{Y}^{KLM,\perp}_{1}
,\mathcal{Y}^{KLM,\perp}_{2}\right)$ of the space of the symmetric HH functions with $(K,L)=(4,0)$ is obtained straightforwardly from the basis~\eqref{bas}, by applying the Gram-Schmidt algorithm, as
\begin{eqnarray}
\mathcal{Y}^{KLM,\perp}_{1}&=&
3.027 650 354 097 491 67\  \mathcal{Y}^{KLM,\mathcal{S}}_{[K^{(1)}]},\\\nonumber
\mathcal{Y}^{KLM,\perp}_{2}&=&
-1.791 182 105 017 056 42\,
\mathcal{Y}^{KLM,\mathcal{S}}_{[K^{(1)}]}\\
&&+6.283 908 019 696 023 53\, \mathcal{Y}^{KLM,\mathcal{S}}_{[K^{(2)}]}.
\end{eqnarray}
\subsection{Identical-boson systems}
In this section, we apply the method discussed in \Sec{secorth}
to the determination of set of linearly independent symmetric HH functions with $L^\pi=0^+$ for systems of four or five identical bosons and for different values of $K$ and $l_{\rm max}$. These symmetric HH functions can be used for studying helium clusters of four and five atoms using soft-core potentials, as done in \Ref{GKV11b} with an unsymmetrized HH basis. 

The numbers of  linearly independent symmetric HH for a four-identical-boson system with $L^\pi=0^+$ are given in \Tab{tab4bos1}.
\begin{table}[ht]
\begin{center}
\begin{tabular}{c c c c c}
\toprule
$K$\textbackslash$l_{\rm max}$&0&2&4&6\\
\midrule
0  &\bf 1 \\    		
2  &\bf 0 \\    		
4  &\bf 2 \\    		
6  &\bf 3 \\    		
8  & 4  &\bf 5 \\		
10 & 5  &\bf 7	\\	
12 & 6  & 10 &\bf 14  \\	
14 & 7  & 12 &\bf 16  \\	
16 & 8  & 14 &\bf 27  \\	
18 & 9  & 16 & 34 &\bf 36 \\
20 & 10 & 18 & 41 &\bf 49 \\
22 & 11 & 20 & 47 & 62 \\
24 & 12 & 22 & 53 & 74 \\
26 & 13 & 24 & 59 & 85 \\
28 & 14 & 26 & 65 & 96 \\
30 & 15 & 28 & 71 & 106\\
\bottomrule
\end{tabular}
\caption{Number of linearly independent symmetric states for a four-identical-boson system with $L^\pi=0^+$ as a function of $l_{\rm max}$. The number is written in bold if, for a given $K$, increasing $l_{\rm max}$ has no impact on the obtained number of linearly independent symmetric states.
}
\label{tab4bos1}
\end{center}
\end{table}
Up to $K=20$, considering $l_{\rm max}=6$ is sufficient to build a full basis of symmetric HH functions. Beyond $K=20$, it is needed to use higher values of $l_{\rm max}$ or, possibly, to consider simultaneously HH functions based on another set of Jacobi coordinates to obtain a full basis of symmetric HH. However, as already mentioned, the contribution of HH with $l_{\rm max}\ge 8$ should be small in bound-state calculations. 
 
For each value of $l_{\rm max}$, for values of $K$ big enough, we have noticed empirically that the number of linearly independent symmetric states increases linearly. This enable us to provide asymptotic relations, given in \Tab{tab4bos2}, between the number of linearly independent states and the grand angular momentum $K$ for given values of $l_{\rm max}$. 
These relations have been checked explicitly up to $K=60$.
\begin{table}[ht]
\begin{center}
\renewcommand{\arraystretch}{1.2}
\begin{tabular}{c c r}
\toprule
$l_{\rm max}$&$K_{\rm as}$&\multicolumn{1}{c}{$N_{\mathcal{S}}$}\\
\midrule
0&4&$K$/2\\
2&12&$K$-2\\
4&20&3$K$-19\\
6&28&5$K$-44\\
\bottomrule
\end{tabular}
\caption{Number of linearly independent symmetric states for a four-identical-boson system with $L^\pi=0^+$ for $K\ge K_{\rm as}$.}
\label{tab4bos2}
\end{center}
\end{table}    

The numbers of  linearly independent symmetric HH for a five-identical-boson system with $L^\pi=0^+$ are given in \Tab{tab5bos1}.
\begin{table}[ht]
\begin{center}
\begin{tabular}{c c c c }
\toprule
$K$\textbackslash$l_{\rm max}$&0&2&4\\
\midrule
0  &\bf 1 \\    
2  &\bf 0 \\
4  &\bf 2 \\
6  &\bf 4 \\
8  &\bf 9 \\
10 & 14 &\bf 18\\
12 & 20 & 36  &\bf 39 \\
14 & 27 & 58  &\bf 69 \\
16 & 35 & 85  &\bf 132\\
18 & 44 & 114 &\bf 230\\
20 & 54 & 147 & 354\\
22 & 65 & 184 & 479\\
24 & 77 & 225 & 619\\
\bottomrule
\end{tabular}
\caption{Number of linearly independent symmetric states for a five-identical-boson system with $L^\pi=0^+$ as a function of $l_{\rm max}$. The number is written in bold if, for a given $K$, increasing $l_{\rm max}$ has no impact on the obtained number of linearly independent symmetric states.}
\label{tab5bos1}
\end{center}
\end{table}
Up to $K=18$, a full basis of symmetric HH functions is obtained by considering HH functions with $l_{\rm sum}\le l_{\rm max}=4$. As in the case of four bosons, for each value of $l_{\rm max}$, some simple asymptotic relations between the number of linearly independent states and the grand angular momentum $K$ have been obtained empirically. However, they are not linear anymore but quadratic, as shown in \Tab{tab5bos2}. This property illustrates the fast increase of the size of the  symmetric HH basis with $K$ when the number of particles becomes high. 
\begin{table}[ht]
\begin{center}
\renewcommand{\arraystretch}{1.2}
\begin{tabular}{c c r}
\toprule
$l_{\rm max}$&$K_{\rm as}$&\multicolumn{1}{c}{$N_{\mathcal{S}}$}\\
\midrule
0&8&$K^2$/8-$K$/4-1\\
2&16&$K^2$/2-5$K$/2-3\\
4&24&2$K^2$-47$K$/2+31\\
\bottomrule
\end{tabular}
\caption{Number of linearly independent symmetric states for a five-identical-boson system with $L^\pi=0^+$ for $K\ge K_{\rm as}$.}
\label{tab5bos2}
\end{center}
\end{table}
%
%
\FloatBarrier
\subsection{Few-nucleon systems}
The methods developed in this paper are applied in this section to the study of few-nucleon systems. First, let us consider the case of the alpha particle. Its ground state is characterized by a total angular momentum $J=0$, a positive parity, and, within a good approximation, an isospin $T=0$. When a realistic nucleon-nucleon potential, which includes a tensor term, is considered, the ground-state wave function contains components with $(L,S)=(0,0)$, $(1,1)$, and $(2,2)$. For these quantum numbers, sets of linearly independent antisymmetric HH functions have been computed. Their size are given in \Tabs{tab4N1}, \ref{tab4N2}, and \ref{tab4N3}.     
\begin{table}[ht]
\begin{center}
\begin{tabular}{c c c c c}
\toprule
$K$\textbackslash$l_{\rm max}$&0&2&4&6\\
\midrule
0&{\bf 1}\\
2&{\bf 1}\\
4&{\bf 4}\\
6&6&{\bf 8}\\
8&8&{\bf 14}\\
10&10&{\bf 24}\\
12&12&34&{\bf 41}\\
14&14&42&{\bf 59}\\
16&16&50&{\bf 90}\\
18&18&58&112&{\bf 128}\\
20&20&66&134&{\bf 176}\\
22&22&74&154&235\\
24&24&82&174&282\\
26&26&90&194&324\\
28&28&98&214&366\\
30&30&106&234&406\\
\bottomrule
\end{tabular}
\caption{Number of linearly independent antisymmetric states for a four-nucleon system with $(L,S,T)^\pi=(0,0,0)^+$ as a function of $l_{\rm max}$. The number is written in bold if, for a given $K$, increasing $l_{\rm max}$ has no impact on the obtained number of linearly independent antisymmetric states.}
\label{tab4N1}
\end{center}
\end{table}
\begin{table}[ht]
\begin{center}
\begin{tabular}{c c c c}
\toprule
$K$\textbackslash$l_{\rm max}$&2&4&6\\
\midrule
2&{\bf 1}\\
4&{\bf 4}\\
6&{\bf 12}\\
8&23&{\bf 27}\\
10&32&{\bf 54}\\
12&41&{\bf 96}\\
14&50&143&{\bf 160}\\
16&59&181&{\bf 250}\\
18&68&217&{\bf 375}\\
20&77&253&488\\
22&86&289&585\\
24&95&325&677\\
26&104&361&767\\
28&113&397&855\\
30&122&433&947\\
\bottomrule
\end{tabular}
\caption{Number of linearly independent antisymmetric states for a four-nucleon system with $(L,S,T)^\pi=(1,1,0)^+$ as a function of $l_{\rm max}$. The number is written in bold if, for a given $K$, increasing $l_{\rm max}$ has no impact on the obtained number of linearly independent antisymmetric states.}
\label{tab4N2}
\end{center}
\end{table}
\begin{table}[ht]
\begin{center}
\begin{tabular}{c c c c}
\toprule
$K$\textbackslash$l_{\rm max}$&2&4&6\\
\midrule
2&{\bf 1}\\
4&{\bf 3}\\
6&{\bf 9}\\
8&16&{\bf 18}\\
10&22&{\bf 36}\\
12&28&{\bf 63}\\
14&34&97&{\bf 102}\\
16&40&122&{\bf 158}\\
18&46&146&{\bf 236}\\
20&52&170&321\\
22&58&194&385\\
24&64&218&445\\
26&70&242&504\\
28&76&266&563\\
30&82&290&622\\
\bottomrule
\end{tabular}
\caption{Number of linearly independent antisymmetric states for a four-nucleon system with $(L,S,T)^\pi=(2,2,0)^+$ as a function of $l_{\rm max}$. The number is written in bold if, for a given $K$, increasing $l_{\rm max}$ has no impact on the obtained number of linearly independent antisymmetric states.}
\label{tab4N3}
\end{center}
\end{table}
For $l_{\rm max}=6$ and up to $K=24$, the numbers of linearly independent antisymmetric HH states have already been given in \Ref{Vi98}. They are in agreement with the ones given in \Tabs{tab4N1}, \ref{tab4N2}, and \ref{tab4N3} except for $(L,S)=(1,1)$ and $K=24$ where a typo in \Ref{Vi98} is suspected.
However, let us stress that, in contrast with \Ref{Vi98}, the sets of linearly independent antisymmetric HH states are here obtained without computing any transformation coefficients, which makes the method much faster. 

As in the case of a four-boson system, for each value of $l_{\rm max}$, the number of linearly independent properly symmetrized HH increases linearly with $K$ (see \Tab{tab4N4}). This linear growing enables one to consider relatively high value of $K$ while keeping the size of the considered HH basis manageable for practical applications, as the accurate computation of the $\alpha$ ground state~\cite{VKR05}. 
\begin{table}[ht]
\begin{center}
\renewcommand{\arraystretch}{1.2}
\begin{tabular}{c c r c r c r}
\toprule
&\multicolumn{2}{c}{$(L,S)=(0,0)$}&\multicolumn{2}{c}{$(L,S)=(1,1)$}&\multicolumn{2}{c}{$(L,S)=(2,2)$}\\
\cmidrule(l){2-3}\cmidrule(l){4-5}\cmidrule(l){6-7}
$l_{\rm max}$&$K_{\rm as}$&\multicolumn{1}{c}{$N_{\mathcal{A}}$}&$K_{\rm as}$&\multicolumn{1}{c}{$N_{\mathcal{A}}$}&$K_{\rm as}$&\multicolumn{1}{c}{$N_{\mathcal{A}}$}\\
0&4&$K$&-&\multicolumn{1}{c}{-}&-&\multicolumn{1}{c}{-}\\
2&12&4$K$-14&8&9$K$/2-13&8&3$K$-8 \\
4&20&10$K$-66&16&18$K$-107&16&12$K$-70\\
6&28&20$K$-194&24&45$K$-403&24&59$K$/2-263\\
\bottomrule
\end{tabular}
\caption{Number of linearly independent antisymmetric states for a four-nucleon system with $T=0$, $\pi=1$ and different $(L,S)$ values for $K\ge K_{\rm as}$.}
\label{tab4N4}
\end{center}
\end{table}

Let us consider now the five-nucleon system $\alpha+N$. This system has no bound state but has two (broad) resonances characterized by a total angular momentum $J=3/2$ or $J=1/2$, a negative parity, and, within a good approximation, an isospin $T=1/2$. For the sake of simplicity, we restrict our discussion to the partial wave $(L,S)=(1,1/2)$, which is dominant for these resonances. The numbers of linearly independent antisymmetric states for the $\alpha+N$ system with $(L,S,T)^\pi=(1,1/2,1/2)^-$ for different values of $K$ and $l_{\rm max}$ are given in \Tab{tab5N1}.
\begin{table}[ht]
\begin{center}
\begin{tabular}{c c c c}
\toprule
$K$\textbackslash$l_{\rm max}$&1&3&5\\
\midrule
1&\bf 1&&\\
3&\bf 7&&\\
5&\bf 36&&\\
7&120&\bf 134&\\
9&279&\bf 399&\\
11&474&\bf 1025 &\\
13&718&2339&\bf 2349\\
15&1011&4083&{\bf 4925}\\
17&1353&6059&{\bf 9617}\\
\bottomrule
\end{tabular}
\caption{Number of linearly independent antisymmetric states for a five-nucleon system with $(L,S,T)^\pi=(1,1/2,1/2)^-$ as a function of $l_{\rm max}$. The number is written in bold if, for a given $K$, increasing $l_{\rm max}$ has no impact on the obtained number of linearly independent antisymmetric states.}
\label{tab5N1}
\end{center}
\end{table}
Up to $K=17$, a full basis of antisymmetric HH functions can be obtained with $l_{\rm max}=5$. As in the case of a five-boson system, for each value of $l_{\rm max}$, the number of linearly independent properly symmetrized HH increases asymptotically in a quadratic way with $K$ (see \Tab{tab5N2}). 
\begin{table}[ht]
\begin{center}
\renewcommand{\arraystretch}{1.2}
\begin{tabular}{c c r}
\toprule
$l_{\rm max}$&$K_{\rm as}$&\multicolumn{1}{c}{$N_{\mathcal{A}}$}\\
\midrule
1&9&49$K^2$/8-25$K$+63/8\\
3&17&349/8$K^2$-426 $K$+5547/8\\
\bottomrule
\end{tabular}
\caption{Number of linearly independent antisymmetric states $N_{\mathcal{A}}$ for a five-nucleon system with $(L,S,T)^\pi=(1,1/2,1/2)^-$ as a function of $l_{\rm max}$ for $K\ge K_{\rm as}$.}
\label{tab5N2}
\end{center}
\end{table}
Up to $K=13$, there are about $4000$ linearly independent antisymmetric HH with $(L,S,T)^\pi=(1,1/2,1/2)^-$ while up to $K=15$, they are about $9000$ and up to $K=17$ about $18000$! Beyond $K=13$, the number of antisymmetric states becomes quite huge and it becomes hardly doable to include all of them. Some selection process is then needed to take only the states with the most important contribution into account. However, does one need to go beyond $K=13$ to describe properly the $\alpha+N$ scattering? When soft inter-nucleon potentials are used, based on the no-core shell model study of the $\alpha+N$ scattering at relative energies up to about 6 MeV~\cite{HLN13}, the answer for this range of energies is no. Indeed, a fair description of the $\alpha+N$ wave function should be obtained by expanding it into an antisymmetric HH basis truncated at $K\le 13$ provided that $\alpha+N$ cluster states are also included in the expansion. 
\FloatBarrier
\section{Conclusion}\label{sec:conc}
In this paper, a numerical approach to build an orthonormal basis of symmetric or antisymmetric hyperspherical harmonic functions has been presented. In particular, two algorithms for computing the transformation coefficients between hyperspherical harmonics constructed from different sets of Jacobi vectors have been improved and/or extended to systems with an arbitrary number of particles. The computational methods developed here should allow one to broaden the scope of applications of the hyperspherical harmonics. In particular, they will enable us in a near future to describe few-body systems containing up to six particles within the HH  approach. Possible applications are the study of the spectra of helium clusters using soft-core potentials, of the properties of the ${^6}{\rm Li}$ nucleus, or of the $\alpha+N$ and $\alpha+d$ scattering. Works towards these directions are on progress and are promising.
\section*{Acknowledgements}
Parts of this work have been done when one of the author (J. D.-E.) was at the INFN-Pisa, first as a postdoctoral fellow and then as a visitor. J. D.-E. thanks the Fonds de la
Recherche Scientifique - FNRS for its financial support during his stay in April-May 2019. 
This work was supported by the Fonds de la Recherche Scientifique - FNRS under Grant Number 4.45.10.08.
\appendix
\section{}
For the sake of completeness, we briefly discuss in this appendix the computation of the integrals involved in recurrence relations~\eqref{TCrec}. More details can be found in \Ref{Vi98}. From the definition~\eqref{HHcoupl} of the HH, these integrals are explicitly given by
\begin{eqnarray}\nonumber
\int d\Omega
\left[\displaystyle\mathcal{Y}^{K+2LM}_{[l'_N L'_{N-1} n'_N]}(\Omega)\right]^* \dfrac{\bm{x}_k\cdot\bm{x}_q}{\rho^2}
\mathcal{Y}^{KLM}_{[l''_N L''_{N-1} n''_N]}(\Omega)
=
\dfrac{1}{4^{N-1}}
\\\nonumber
\times\left(\prod^N_{j=2}
\mathcal{N}^{\alpha_{K'_{j-1}},\beta_{l'_j}}_{n'_j}\mathcal{N}^{\alpha_{K''_{j-1}},\beta_{l''_j}}_{n''_j}
\right)\int_{4\pi} d\hat{x}_1\ldots \int_{4\pi} d\hat{x}_N 
\int^1_{-1} dz_2 \ldots \int^1_{-1} dz_N\\\nonumber
\times\Big[\Big[\ldots\Big[Y_{l'_1}(\hat{x}_1)Y_{l'_2}(\hat{x}_2)\Big]_{L'_2} \ldots Y_{l'_{N-1}}(\hat{x}_{N-1})\Big]_{L'_{N-1}} Y_{l'_N}(\hat{x}_N)\Big]^*_{LM}\\\nonumber
\times\Big[\Big[\ldots\Big[Y_{l''_1}(\hat{x}_1)Y_{l''_2}(\hat{x}_2)\Big]_{L''_2} \ldots Y_{l''_{N-1}}(\hat{x}_{N-1})\Big]_{L''_{N-1}} Y_{l''_N}(\hat{x}_N)\Big]_{LM}\\\nonumber
\times
\dfrac{\bm{x}_k\cdot\bm{x}_q}{\rho^2}
\prod^N_{j=2}\left(\dfrac{1+z_j}{2}\right)^{(l'_j+l''_j+1)/2}\left(\dfrac{1-z_j}{2}\right)^{(K'_{j-1}+K''_{j-1}+3j-5)/2}
\\\label{intrec}
\times
P^{\alpha_{K'_{j-1}},\beta_{l'_j}}_{n'_j}(z_j)P^{\alpha_{K''_{j-1}},\beta_{l''_j}}_{n''_j}(z_j),
\end{eqnarray}
where $k,q=1,\ldots,N$ and the substitutions $z_j=\cos 2\phi_j$ for $j=2,\ldots,N$ have been performed. The scalar product of Jacobi coordinates divided by $\rho^2$ can be written for $k,q=1,\ldots,N$ as
\begin{align}\nonumber
\dfrac{\bm{x}_k\cdot\bm{x}_q}{\rho^2}&= \dfrac{x^2_k}{\rho^2}&{\rm if\ }k=q\\\label{prod1}
&=-\dfrac{4\pi}{\sqrt{3}} \dfrac{x_k x_q}{\rho^2}
\left[Y_1(\hat{x}_k)Y_1(\hat{x}_q)\right]_{00}&{\rm if\ }k\ne q,
\end{align}
with
\begin{eqnarray}\label{prod2}
\dfrac{x_1}{\rho}&=&\prod^N_{j=1} \left(\dfrac{1-z_j}{2}\right)^{1/2}
\end{eqnarray}
and
\begin{eqnarray}\label{prod3}
\dfrac{x_l}{\rho}&=&\left(\dfrac{1+z_l}{2}\right)^{1/2}\prod^N_{j=l+1} \left(\dfrac{1-z_j}{2}\right)^{1/2}
\end{eqnarray}
for $l=2,\ldots,N$. For $k=q$, the angular integrals in \Eq{intrec} reduce to a product of Kronecker delta: $\delta_{l'_1 l''_1}\ldots \delta_{l'_N l''_N}
\delta_{L'_2 L''_2}\ldots \delta_{L'_N L''_N}$. For $k\neq q$, they can be expressed, by means of the Wigner-Eckaert theorem and of the standard techniques for computing reduced matrix elements (see for instance \Ref{Ed57}), as a product of Wigner $3j$ and $6j$ symbols and of the square roots of a ratio of integer numbers. Many of these angular integrals vanish. 
Taking \Eqs{prod1}-\eqref{prod3} into account, the integrals over $z_2,\ldots,z_N$ in \Eq{intrec} reduce to a product of one-dimensional integrals. Each of them can be written under the form
\beq\label{intJacobi}
\int^{1}_{-1} dz\, (1-z)^{a} (1+z)^{b} \mathcal{P}_m(z),
\eeq
where $a,b=0$ or $1/2$ and $\mathcal{P}_m$ is a polynomial of degree $m\leq K+5N/2-2$. Integrals of the form~\eqref{intJacobi} can be computed exactly, up to rounding errors, by appropriate Gauss-Jacobi quadratures with $N_G\geq (m+1)/2$ integration points. Some of them vanish.

\newcommand{\bibGy}{Gy}\newcommand{\bibZs}{Zs}\newcommand{\bibPH}{P.-H}\newcommand{\NPA}{Nucl.
  Phys. A~}\newcommand{\PRA}{Phys. Rev. A~}\newcommand{\PRC}{Phys. Rev.
  C~}\newcommand{\JPAMG}{J. Phys. A~}\newcommand{\JPAMT}{J. Phys.
  A~}\newcommand{\JPBAMO}{J. Phys. B~}\newcommand{\FBS}{Few-{B}ody
  {S}yst.~}\newcommand{\SIAMJMAA}{SIAM J. Matrix Anal.
  A.~}\newcommand{\EPJp}{Eur. Phys. J. Plus~}\newcommand{\PLB}{Phys. Lett.
  B~}\newcommand{\JPG}{J. Phys. G~}\newcommand{\PPNP}{Prog. Part. Nucl.
  Phys.~}\newcommand{\PTEP}{Prog. Theor. Exp. Phys.~}





\begin{thebibliography}{10}
\expandafter\ifx\csname url\endcsname\relax
  \def\url#1{\texttt{#1}}\fi
\expandafter\ifx\csname urlprefix\endcsname\relax\def\urlprefix{URL }\fi
\expandafter\ifx\csname href\endcsname\relax
  \def\href#1#2{#2} \def\path#1{#1}\fi

\bibitem{LO13}
W.~Leidemann, G.~Orlandini, Modern ab initio approaches and applications in
  few-nucleon physics with ${A}\geq4$, \PPNP 68 (2013) 158--214.

\bibitem{MS15}
C.~Meyer, E.~Swanson, Hybrid mesons, \PPNP 82 (2015) 21--58.

\bibitem{RVV18}
J.-M. Richard, A.~Valcarce, J.~Vijande, Few-body quark dynamics for doubly
  heavy baryons and tetraquarks, Phys. Rev. C 97 (2018) 035211.

\bibitem{Pl15}
W.~Plessas, The constituent-quark model --- {N}owadays, Int. J. Mod. Phys. A 30
  (2015) 1530013.

\bibitem{ZB35}
F.~Zernike, H.~Brinkman, {H}yperspharische {F}unktionen und die in spharischen
  {B}ereichen orthogonalen {P}olynome, Proc. K. Ned. Akad. Wet. 38 (1935)
  161--170.

\bibitem{Av89}
J.~Avery, {H}yperspherical {H}armonics, Kluwer, 1989.

\bibitem{KRV08}
A.~Kievsky, S.~Rosati, M.~Viviani, L.~E. Marcucci, L.~Girlanda, A
  high-precision variational approach to three- and four-nucleon bound and
  zero-energy scattering states, \JPG 35 (2008) 063101.

\bibitem{BN97}
N.~Barnea, A.~Novoselsky, Construction of hyperspherical functions symmetrized
  with respect to the orthogonal and the symmetric groups, Ann. Phys. 256
  (1997) 192--225.

\bibitem{BN98}
N.~Barnea, A.~Novoselsky, Hyperspherical wave functions with orthogonal and
  permutational symmetry, Phys. Rev. A 57 (1998) 48--58.

\bibitem{Ba99}
N.~Barnea, Hyperspherical functions with arbitrary permutational symmetry:
  Reverse construction, Phys. Rev. A 59 (1999) 1135--1146.

\bibitem{BLO00}
N.~Barnea, W.~Leidemann, G.~Orlandini, State dependent effective interaction
  for the hyperspherical formalism, Phys. Rev. C 61 (2000) 054001.

\bibitem{BLO03}
N.~Barnea, W.~Leidemann, G.~Orlandini, Improved effective interaction for the
  hyperspherical formalism, Phys. Rev. C 67 (2003) 054003.

\bibitem{BLO10}
N.~Barnea, W.~Leidemann, G.~Orlandini, Hyperspherical effective interaction for
  nonlocal potentials, Phys. Rev. C 81 (2010) 064001.

\bibitem{BBS12}
S.~Bacca, N.~Barnea, A.~Schwenk, Matter and charge radius of ${}^{6}${H}e in
  the hyperspherical-harmonics approach, Phys. Rev. C 86 (2012) 034321.

\bibitem{Ti02}
N.~K. Timofeyuk, Shell model approach to construction of a hyperspherical basis
  for a identical particles: Application to hydrogen and helium isotopes, Phys.
  Rev. C 65 (2002) 064306.

\bibitem{Ti04}
N.~K. Timofeyuk, Hyperspherical harmonics with orthogonal symmetry in the shell
  model approach and its application to light nuclei, Phys. Rev. C 69 (2004)
  034336.

\bibitem{Ti08}
N.~K. Timofeyuk, Improved procedure to construct a hyperspherical basis for the
  ${N}$-body problem: Application to bosonic systems, Phys. Rev. C 78 (2008)
  054314.

\bibitem{RSI11}
S.~T. Rittenhouse, J.~von Stecher, J.~P. D'Incao, N.~P. Mehta, C.~H. Greene,
  The hyperspherical four-fermion problem, \JPBAMO 44 (2011) 172001.

\bibitem{RB12}
D.~Rakshit, D.~Blume, Hyperspherical explicitly correlated gaussian approach
  for few-body systems with finite angular momentum, Phys. Rev. A 86 (2012)
  062513.

\bibitem{DG14}
K.~M. Daily, C.~H. Greene, Extension of the correlated gaussian hyperspherical
  method to more particles and dimensions, Phys. Rev. A 89 (2014) 012503.

\bibitem{Su15}
Y.~Suzuki, Adiabatic hyperspherical approach to large-scale nuclear dynamics,
  \PTEP 2015 (2015) 043D05.

\bibitem{SV19}
Y.~Suzuki, K.~Varga, Constrained correlated-gaussians for hyperspherical
  calculations, \FBS 60 (2019) 3.

\bibitem{GKV09}
M.~Gattobigio, A.~Kievsky, M.~Viviani, P.~Barletta, Harmonic hyperspherical
  basis for identical particles without permutational symmetry, Phys. Rev. A 79
  (2009) 032513.

\bibitem{GKV11a}
M.~Gattobigio, A.~Kievsky, M.~Viviani, Nonsymmetrized hyperspherical harmonic
  basis for an ${A}$-body system, Phys. Rev. C 83 (2011) 024001.

\bibitem{DBL13}
S.~Deflorian, N.~Barnea, W.~Leidemann, G.~Orlandini, Nonsymmetrized
  hyperspherical harmonics with realistic nn potentials, \FBS 54 (2013)
  1879--1887.

\bibitem{RR70}
J.~Raynal, J.~Revai, Transformation coefficients in the hyperspherical approach
  to the three-body problem, Il Nuovo Cimento 68 (1970) 612--622.

\bibitem{JKS77}
R.~I. Jibuti, N.~B. Krupennikova, N.~I. Shubitidze, On the four-body problem,
  Theor. Math. Phys. 32 (1977) 704--713, translated from {T}eoreticheskaya i
  {M}atematicheskaya {F}izika 32 (1977) 223--236.

\bibitem{Vi98}
M.~Viviani, Transformation coefficients of hyperspherical harmonic functions of
  an ${A}$-body system, Few-{B}ody {S}yst. 25 (1998) 177--187.

\bibitem{VKR05}
M.~Viviani, A.~Kievsky, S.~Rosati, Calculation of the
  $\ensuremath{\alpha}$-particle ground state within the hyperspherical
  harmonic basis, Phys. Rev. C 71 (2005) 024006.

\bibitem{Ef95}
V.~Efros, Hyperspherical/oscillator brackets and symmetrized states: A simple
  algorithm for ${A} > 3$ few-body calculations, \FBS 19 (1995) 167--173.

\bibitem{GL96}
G.~H. Golub, C.~F. {V}an {L}oan, Matrix Computations, 3rd Edition, Johns
  Hopkins, Baltimore, 1996.

\bibitem{AS65}
M.~Abramowitz, I.~A. Stegun, Handbook of Mathematical Functions, Dover, New
  York, 1965.

\bibitem{NumRecFor}
W.~H. Press, S.~A. Teukolsky, W.~T. Vetterling, B.~P. Flannery, Numerical
  Recipes in Fortran 77: The Art of Scientific Computing, 2nd Edition,
  Cambridge, New York, 1992.

\bibitem{GKV11b}
M.~Gattobigio, A.~Kievsky, M.~Viviani, Spectra of helium clusters with up to
  six atoms using soft-core potentials, Phys. Rev. A 84 (2011) 052503.

\bibitem{HLN13}
G.~Hupin, J.~Langhammer, P.~Navr\'atil, S.~Quaglioni, A.~Calci, R.~Roth, Ab
  initio many-body calculations of nucleon-${{^4}{\rm He}}$ scattering with
  three-nucleon forces, Phys. Rev. C 88 (2013) 054622.

\bibitem{Ed57}
A.~R. Edmonds, Angular {M}omentum in {Q}uantum {M}echanics, {P}rinceton
  {U}niversity, Princeton, 1957.

\end{thebibliography}





\end{document}